# Combining Bioimpedance and EMG Measurements for Reliable Muscle Contraction Detection

Roman Kusche*, and Martin Ryschka*

*Abstract— Objective:* **Muscle contractions are commonly detected by performing EMG measurements. The major disadvantage of this technique is that mechanical disturbances to the electrodes are in the same frequency and magnitude range as the desired signal. In this work we propose an approach and a realized measurement system to combine EMG and bioimpedance measurements for higher reliabilities of muscle contraction detections.** *Methods:* **We propose the development of a modular four-channel measurement system, whereat each channel is capable of acquiring EMG, the bioimpedance magnitude and phase, simultaneously. The modules are synchronized by an additional interface board, which communicates with a PC. A graphical user interface enables to control the bioimpedance excitation current in a range from 100 µA to 1 mA in a frequency range from 50 kHz to 333 kHz.** *Results:* **A system characterization demonstrated that bioimpedance magnitude changes of less than 250 ppm and phase changes below 0.05° can be detected reliably. Measurements from a subject have shown the timing relationship between EMG and bioimpedance signals as well as their robustness against mechanical disturbances. A measurement of five exemplary hand gestures has demonstrated the increase of usable information for detecting muscle contractions.** *Conclusion:* **Bioimpedance measurements of muscles provide useful information about contractions. Furthermore, the usage of a known high-frequency excitation current enables a reliable differentiation between the actual information and disturbances.** *Significance:* **By combining EMG and bioimpedance measurements, muscle contractions can be detected much more reliably. This setup can be adopted to prostheses and many other human-computer interfaces.**

## I. Introduction

In the past several years the detection of muscle contractions has become an important technique in biomedical engineering. It can either be acquired at a single position of the body to analyze the state of a single muscle (group) or it is acquired at different positions simultaneously. The gathered information is commonly used to control computer systems, robots, exoskeletons or prostheses [1], [2], [3]. Especially for the last example, a high reliability is important to avoid misinterpretations of the signals, which could cause dangerous situations to the user.

The commonly used measurement technique to detect muscle contractions is the Electromyography (EMG) [4]. This method is based on acquiring the action potentials of the contracted muscles [5]. Typically, these signals result in differential voltages in millivolt ranges at the body surface and can be acquired using surface electrodes [6]. Since the most dominant signal frequency range is from 10 Hz up to 100 Hz [7], the electrical measurement instrumentation can be very elementary. Further methods to detect muscle contractions are the mechanomyography (MMG) [8] and optical setups [9]. Algorithms to combine a set of several signals to get information about movements are often machine-learning based [10].

One of the most challenging problems are motion artifacts, since the fixation of optical, ultrasound or acceleration sensors to the human body is problematic. But also the most commonly acquired EMG signal is very sensitive to motion artifacts [11]. Especially, forces to the electrode-skin interfaces result in significant signal disturbances [12]. Since these disturbances are in a similar frequency range like the actual EMG signal, it is difficult to differentiate between both of them.

A promising measurement technique which might be more robust towards these disturbances is the electrical impedance myography (EIM). This technique is based on measuring the changes of the muscles' electrical bioimpedance, which occur during muscle contractions, caused by geometrical changes [13], [14]. The major advantage of this measurement method is that muscle contraction information is modulated onto an excitation current of much higher frequency than that of motion artifacts. Therefore, the subsequent filtering of motion artifacts is very simple. Obviously, it is also advantageous to extend the EMG information by the two additional dimensions of information, the magnitude and the phase of the bioimpedance (BioZ).

In this work, we propose a measurement method to acquire

Manuscript received xxx xx, 201x; revised xxx xx, 201x and xxx xx, 201x; accepted xxx xx, 201x. Date of publication xxx xx, 201x; date of current version xxx xx, 201x. This work was supported by the German Federal Ministry of Education and Research (BMBF) under the project INOPRO (FKZ16SV7666).

*R. Kusche and M. Ryschka are with the Laboratory of Medical Electronics (LME), Luebeck University of Applied Sciences, 23562 Luebeck, Germany. (correspondence email: roman.kusche@th-luebeck.de).

Digital Object Identifier xx.xxxx/xxxxxx

EMG signals as well as bioimpedance signals simultaneously from the human body to increase the reliability of muscle contraction detections. After explaining the measurement approach, we describe the development of a four-channel measurement system, capable of acquiring the EMG signal and the magnitude as well as the phase of the bioimpedance at four different positions of the human body. The developed system is focused, but not limited to the usage in combination with upper limb prostheses.

After a system characterization, measurements from a human subject demonstrate the behavior of EIM signals in comparison to EMG signals. Additionally, the robustness towards motion artifacts of both the methods is shown. Finally, all four channels are used to measure five common hand gestures and to compare the resulting signals. For comparison purposes, this measurement is repeated for two isometric hand gestures.

## II. Measurement Approach

Although, the EMG as well as the EIM signals contain information about the muscle state, the origins of both the signals are different. The EMG signal, acquired by means of surface electrodes, is the summation of single fiber potentials of a muscle [15]. Activating the muscle results in measureable voltage amplitudes of up to 2 mV [5] in the frequency range below 100 Hz [7].

The EIM is based on the fact, that geometrical changes during muscle contractions cause changes in the electrical impedance of the observed muscle. Since the muscle is a very anisotropic conductor, this effect depends on the positioning of the electrodes [14].

The major difference between both the signals is, that EMG detects muscle activations, whereas EIM detects actual muscle contractions.

The approach of measuring both signals simultaneously is depicted in Fig. 1, in which the EMG signal is modelled as a voltage source $V_{EMG}$ and the bioimpedance is represented by an impedance $Z_{Bio}$. Via the electrodes $E_1$ and $E_2$, a small known AC current $I_{BioZ}$ with a frequency in kHz range is applied to the tissue under test. This current flows through the EMG voltage source and the bioimpedance. Two additional electrodes ($E_3$ and $E_4$) are used do acquire the occurring voltage drop. Because of the significant influence of the electrode-skin interfaces [16], the electrodes are modelled as serial circuits consisting of electrode-skin impedances ($Z_{ESI}$) and DC voltage sources ($V_{HC}$), representing the occurring half-cell voltages. Assuming the input impedance of the differential amplifiers to be infinity, this model leads to the actual measurable addition of voltages with different frequencies, as shown in equation 1.

$$V_{Meas.} = V_{EMG} + Z_{Bio} \cdot I_{BioZ} + V_{HC,V+} - V_{HC,V-} \quad (1)$$

Two different paths of signal processing, each beginning with amplification, are applied for the EMG and bioimpedance signal extraction, respectively.

For obtaining the EMG signal and removing high frequency disturbances as generated by the EIM, a band pass filter

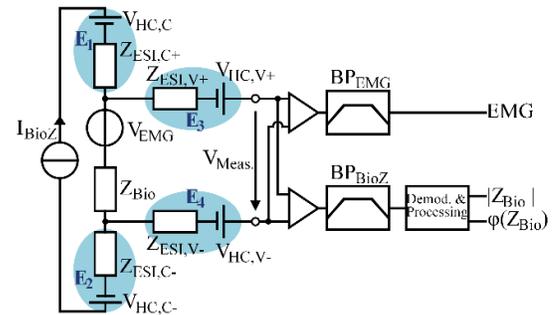

Fig. 1. Illustration of the simplified measurement approach. The blue ellipses represent the electrical behavior of the electrode-skin interfaces. The current source $I_{BioZ}$ is part of the electrical circuitry, whereas the voltage source $V_{EMG}$ symbolizes the EMG signal source of the muscle.

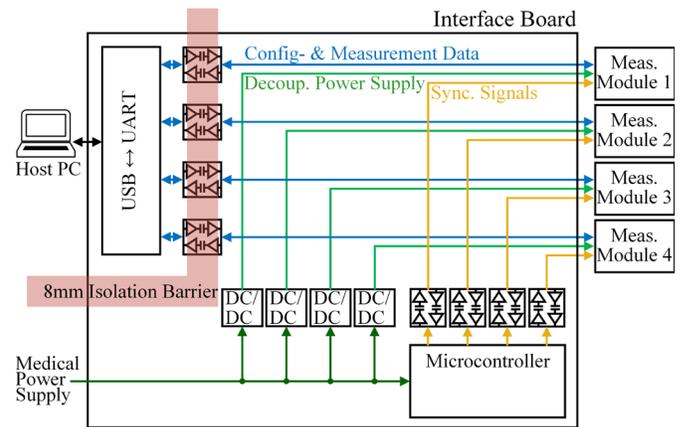

Fig. 2. Block diagram of the whole developed measurement system with a focus on the interface board, which connects a host PC with the four measurement modules. The red bar illustrates the galvanic isolation barrier of 8 mm width to fulfill the IEC60601-1. The blue arrows indicate the exchange of data and the green arrows represent the power supply connections, which decouple the modules electrically from each other. For the same reason, the yellow synchronization connections are interrupted by digital isolators as well.

($BP_{EMG}$) is used. When $V_{HC,V+}$ and $V_{HC,V-}$ are DC voltages, the band pass filter attenuates also these signal components. However, if mechanical forces, applied to the electrodes, change the half-cell voltages quickly over time, the occurring AC voltages are not removed anymore.

The bioimpedance path includes a band pass filter ($BP_{BioZ}$) for removing undesirable frequency components as well. The commonly used frequency range of bioimpedance measurements is much higher than that of typical motion artifacts [16], [11]. Therefore, the lower cut-off frequency of the filter is much higher than the frequency range of the occurring disturbances. To calculate the bioimpedance magnitude $|Z_{Bio}|$ and phase $\varphi(Z_{Bio})$, the signal is demodulated by an IQ (In-phase and Quadrature)-demodulation in combination with low pass filters. Since this demodulation technique utilizes the knowledge about the excitation frequency of the current source ($I_{BioZ}$), other frequency components are cancelled out.

## III. System Development

To allow a high flexibility regarding the number of channels, a modular four-channel measurement system was designed. Significant requirements were the synchronicity of EMG and EIM measurements as well as the medical electrical safety. Since the device is intended to measure bioimpedances at several points of the subject simultaneously, a galvanic decoupling of all channels had to be implemented [17].

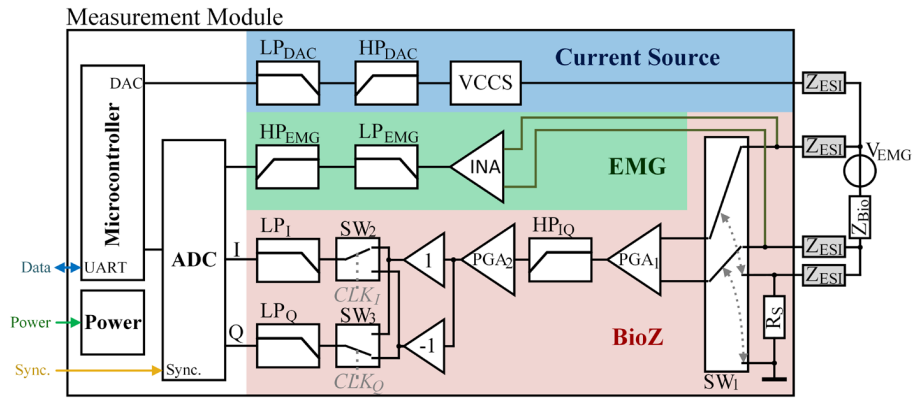

Fig. 3. Detailed block diagram of a measurement module. On the left side there are the connections to the previously described interface board. On the right side the connected measurement subject, which is connected via electrodes, is drawn. The measurement module consists of a microcontroller, a power section, an analog-to-digital converter and three major analog electronics blocks. The blue block marks the current source circuit, the green rectangle illustrates the EMG circuit and the red area indicates the bioimpedance measurement circuit, which is based on an IQ-demodulation.

After an overview of the whole system, the developed measurement modules are explained more detailed. Finally, the realization of the device is presented.

*A. Concept*

In Fig. 2, the block diagram of the developed measurement system is depicted. It consists mainly of an interface board and four measurement modules. The interface board is intended to convert the USB interface of a linked host PC to four UART connections for the measurement modules. In accordance with the safety requirements for medical electrical devices (IEC 60601-1) the measurement modules together with their connection on the interface board are regarded as Applied Parts, which need to be isolated. The isolation for data transport is accomplished by digital isolators (ISO7721, Texas Instruments, Dallas, TX, US) which realize an 8 mm isolation gap. For powering the measurement system, an external 5 $V_{DC}$ medical power supply is used (MPU31-102, SINPRO Electronics, Pingtung City, TW). To decouple the measurement channels from each other, necessary ±5 $V_{DC}$ supply voltages are generated separately via isolated DC/DC converters (MTU2D0505, Murata Power Solutions, Westborough, MA, US).

Synchronizing all boards is realized by a common synchronization clock, which is generated by a microcontroller (ATSAM4S16C, Microchip Technology, Chandler, AZ, US) and transmitted via digital isolators (MAX14930, Maxim Integrated, San Jose, CA, US) to the measurement modules.

*B. Measurement Modules*

Each measurement module is intended to acquire an EMG signal and the complex bioimpedance from a human subject simultaneously. Since the modules are also intended to be used in combination with other devices like prostheses, the data interface is desired to be simple.

The EMG signal characteristic is well known and has already been described above. However, the exact impedance behavior of the muscle during contractions is difficult to predict. In literature, impedance changes in the range from about 1 % to 5 % have been reported [18]. These values can vary widely and depend on the specific measurement setup and the used excitation frequency. Therefore, the EIM circuitry has to be flexible regarding excitation frequency and impedance range. Based on these requirements, the measurement modules were developed.

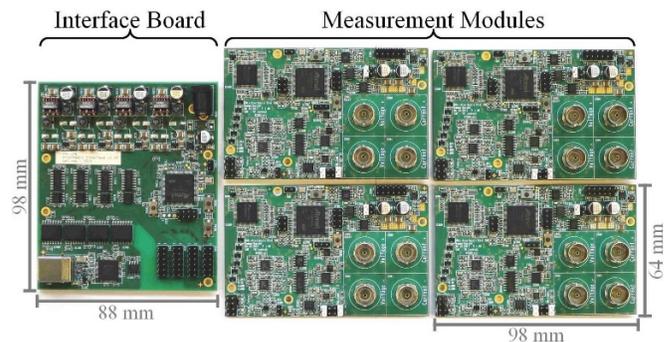

Fig. 4. Photograph of the interface board and four measurement modules. The interface board consists of about 200 components and has dimensions of 98 x 88 mm². On the measurement modules about 300 components are populated and its dimensions are 98 x 64 mm².

In Fig. 3, the block diagram of a measurement module is shown. After receiving the configuration data from the previously described interface board, the microcontroller's (ATSAM4S16C, Microchip Technology, Chandler, AZ, US) digital-to-analog converter (DAC) generates a sinusoidal voltage in the frequency range from 50 kHz to 333 kHz with a sampling rate of 1 MSPS. In this band, the bioimpedance is very sensitive against frequency changes [16]. An active low-pass filter ($LP_{DAC}$, N=4, $f_c$=350 kHz) attenuates the DAC artifacts and a passive high-pass filter ($HP_{DAC}$, N=1, $f_c$=200 Hz) removes the DC component. Utilizing a voltage-controlled current source (VCCS) circuit [17], [19], this AC voltage is converted into an AC current. This current flows via electrodes through the examined bioimpedance ($Z_{Bio}$) and a known shunt resistor ($R_S$).

The occurring voltage drop over the bioimpedance, caused by the excitation current, as well as the superposition of the EMG voltage, is detected by two additional inner electrodes.

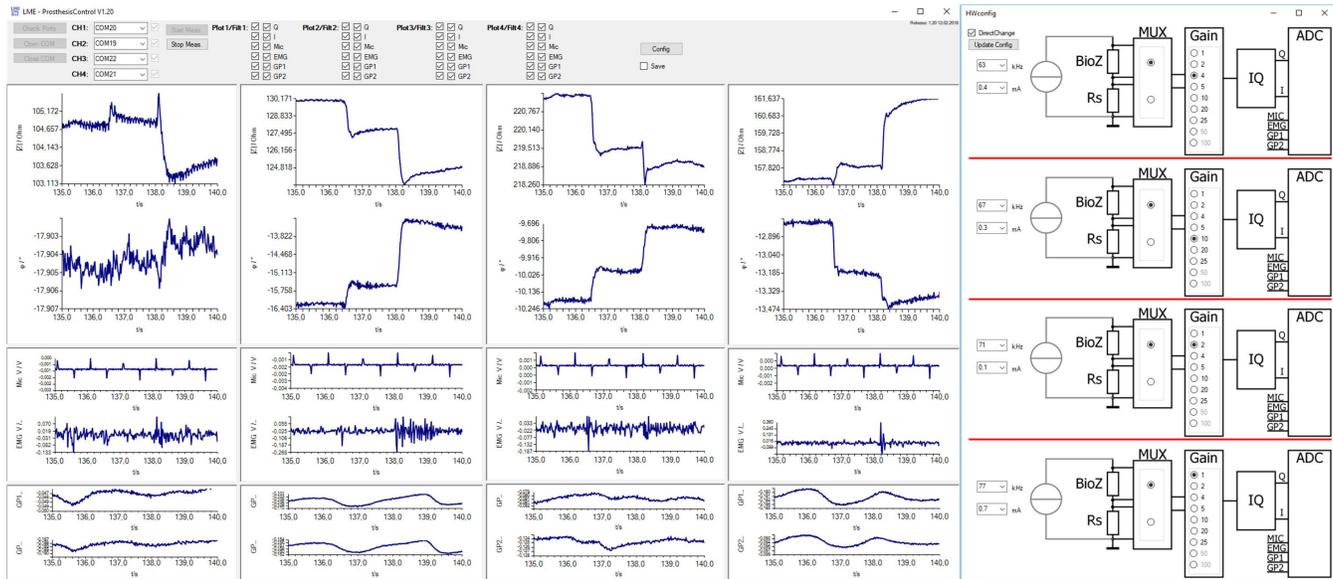

Fig. 5. Screenshot of the GUI to control the measurement system and to display the acquired data in real-time. Each column shows the six digitized signals of a measurement module. On the right side, the configuration window is shown. It can be used to change the frequency and amplitude of the excitation current as well as for configuring the gain of the programmable gain amplifiers.

The processing of this differential voltage signal is split into two different paths. The green colored block indicates the EMG signal processing and the red block the bioimpedance measurement circuitry.

In the green EMG block, the differential voltage is pre-amplified and converted into a single-ended signal by an instrumentational amplifier (INA, INA126, Texas Instruments, Dallas, TX, US, G=11). To remove the high frequency signal components, caused by the bioimpedance measurement, the signal is filtered by an active low-pass filter ($LP_{EMG}$, N=2, $f_c$=100 Hz, G=5). Afterwards, the DC voltage component is removed by a passive high-pass filter ($HP_{EMG}$, N=1, $f_c$=3 Hz) and the signal is transmitted to the analog-to-digital converter (ADC, ADS131E06, Texas Instruments, Dallas, TX, US).

In the red BioZ block, the acquired differential voltage is connected to an analog switch ($SW_1$, ADG1236, Analog Devices, Norwood, MA, US), which can be used to choose between measuring the voltage drop over the bioimpedance or over the shunt resistor to determine the actual measurement current. The selected signal is amplified by a programmable gain amplifier ($PGA_1$, AD8250, Analog Devices, Norwood, MA, US, G={1, 2, 5, 10}) and high-pass-filtered ($HP_{IQ}$, N=1 $f_c$=1 kHz) to attenuate low-frequency noise and the DC voltage, caused by the electrodes' half-cells. Afterwards, the signal is amplified by the $PGA_2$ (AD8250). For the IQ-demodulation, the signal's In-phase (I) and Quadrature (Q) components have to be determined [20]. Therefore, the signal passes two separate paths. In both the paths, analog switches ($SW_2$, $SW_3$, MAX4523, Maxim Integrated, San Jose, CA, US) alternate between the original and the inverted signal, clocked by the microcontroller with the current excitation signal frequency. The clock signals ($CLK_I$, $CLK_Q$) are phase shifted by 90° to realize the IQ-demodulation method [20]. Since only the occurring time dependent DC components of both the switching outputs contains the useful information, the signals are low pass filtered ($LP_I$, $LP_Q$, N=6, $f_c$=1 kHz) before digitization.

The analog processed EMG, the I-component as well as the Q-component signals are digitized synchronously by the 6-channels ADC with a sampling rate of 1000 SPS and a resolution of 24 bits. The remaining three ADC channels are prepared for future use purposes. Via a serial peripheral interface (SPI), the ADC transmits the digitized data to the microcontroller.

### C. Implementation

To realize the interface board as well as the measurement modules, 4-layers printed circuit boards (PCB) have been developed, as shown in the photograph in Fig. 4. To control the device and to display the measured data in real-time, a graphical user interface (GUI) has been programmed in C# language. A screenshot is depicted in Fig. 5. Since it is assumed, that during a measurement the applied excitation current is stable, the GUI automatically measures the current just once at the end of a measurement procedure. For that purpose, it changes the position of $SW_1$ to the shunt resistor.

The excitation frequency of each measurement module can be chosen from 17 discrete values in the range from 50 kHz to 333 kHz. To keep the occurring voltage drops over the bioimpedances in the range of typical EMG signals as well as for electrical safety purposes, the excitation current amplitudes are limited and can be chosen from 11 discrete values in the range from 100 µA to 1 mA. Selectable total PGA gains for the voltage measurement are G=1; 2; 4; 5; 10; 25; 50; 100.

The total power consumption with all four active channels is about 5.5 W, whereas each measurement module consumes about 1 W.

## IV. SYSTEM PERFORMANCE

Before performing measurements on subjects, the system performance has to be analyzed. This includes the determination of measurement errors such as systematic, statistical as well as the system's long-term drift errors. Finally, the synchronicity of all measurement channels is verified.

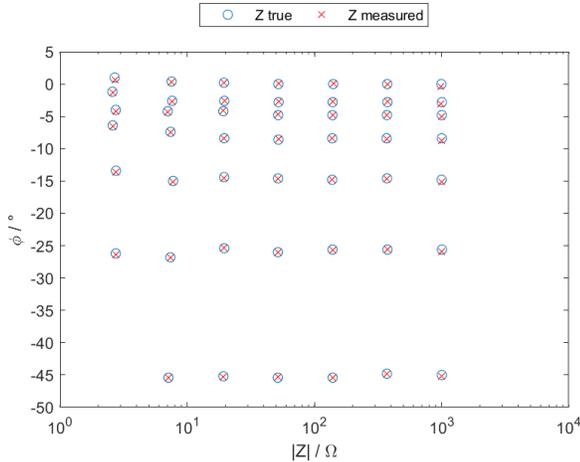

Fig. 6. Plot of 48 actual impedance values (blue circles) and measured values (red crosses). Each value was measured for a duration of 1 s and averaged. The excitation current was chosen to be 1 mA with a frequency of 100 kHz.

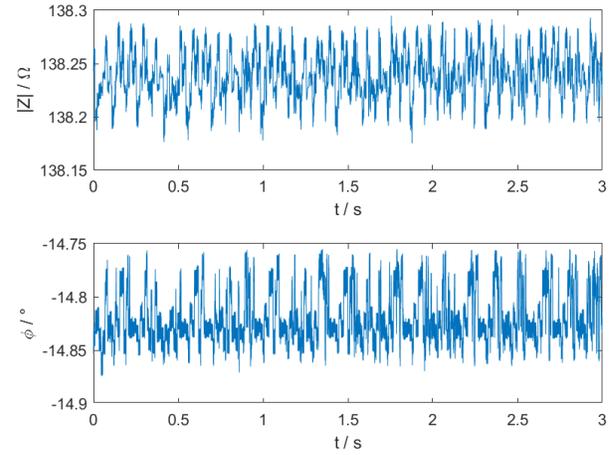

Fig. 7. Exemplary measurement of a complex impedance over 3 s, 3000 samples respectively. The true values are $|Z|=138.1\ \Omega$; $\varphi(Z)=-14.8°$. The standard deviations of this measurement were calculated to be 20 m$\Omega$ and 0.024°.

TABLE I
SYSTEMATIC ERRORS

| | | | | | | | |
|---|---|---|---|---|---|---|---|
| $|Z|_{true}$ / $\Omega$ | 2.7 | 7.3 | 19.5 | 52.1 | 139.6 | 373.6 | 1000 |
| $\varphi_{true}$ | 0° | 0° | 0° | 0° | 0° | 0° | 0° |
| Gain | 25 | 25 | 25 | 25 | 10 | 4 | 1 |
| $|\Delta Z|$ / m$\Omega$ | 4.8 | 10 | 35 | 78 | 177 | 392 | 201 |
| $|\Delta Z|$ / ‰ | 1.8 | 1.4 | 1.8 | 1.5 | 1.3 | 1.0 | 0.2 |
| $|\Delta\varphi|$ / ° | 0.55 | 0.18 | 0.09 | 0.08 | 0.01 | 0.09 | 0.3 |
| $|Z|_{true}$ / $\Omega$ | 139.6 | 139.6 | 139.6 | 139.6 | 139.6 | 139.6 | 139.6 |
| $\varphi_{true}$ | 0° | -2.7° | -4.8° | -8.4° | -14.6° | -25.7° | -45° |
| Gain | 10 | 10 | 10 | 10 | 10 | 10 | 10 |
| $|\Delta Z|$ / m$\Omega$ | 177 | 146 | 158 | 147 | 162 | 109 | 67 |
| $|\Delta Z|$ / ‰ | 1.3 | 1.0 | 1.1 | 1.1 | 1.2 | 0.8 | 0.5 |
| $|\Delta\varphi|$ / ° | 0.01 | 0.07 | 0.03 | 0.00 | 0.02 | 0.11 | 0.01 |

TABLE II
STATISTICAL ERRORS

| | | | | | | | |
|---|---|---|---|---|---|---|---|
| $|Z|_{true}$ / $\Omega$ | 2.7 | 7.3 | 19.5 | 52.1 | 139.6 | 373.6 | 1000 |
| $\varphi_{true}$ | 0° | 0° | 0° | 0° | 0° | 0° | 0° |
| Gain | 25 | 25 | 25 | 25 | 10 | 4 | 1 |
| STD($|Z|$) / m$\Omega$ | 0.55 | 1.2 | 4.1 | 9.5 | 23 | 74 | 202 |
| STD($|Z|$) / ppm | 204 | 164 | 210 | 182 | 165 | 198 | 202 |
| STD($\varphi$) / ° | 0.017 | 0.017 | 0.018 | 0.023 | 0.023 | 0.016 | 0.025 |
| $|Z|_{true}$ / $\Omega$ | 139.6 | 139.6 | 139.6 | 139.6 | 139.6 | 139.6 | 139.6 |
| $\varphi_{true}$ | 0° | -2.7° | -4.8° | -8.4° | -14.6° | -25.7° | -45° |
| Gain | 10 | 10 | 10 | 10 | 10 | 10 | 10 |
| STD($|Z|$) / m$\Omega$ | 23 | 20 | 24 | 22 | 20 | 14 | 12 |
| STD($|Z|$) / ppm | 165 | 143 | 172 | 158 | 143 | 100 | 86 |
| STD($\varphi$) / ° | 0.023 | 0.023 | 0.023 | 0.024 | 0.024 | 0.019 | 0.019 |

### A. Systematic Errors

Since especially the impedance changes over time are of interest, in many bioimpedance measurement applications the systematic error is not as important as the statistical error. Nevertheless, the system's systematic errors were analyzed by measuring known impedances in the range from magnitudes $|Z|=2.7\Omega$ to 1000 $\Omega$ and from phase shifts $\varphi=0°$ to $\varphi=-45°$. The tolerances of the used impedances are 0.1 % and 0.1°, respectively. To eliminate the influence of the statistical errors during this process, each impedance was measured for a duration of 1 s, corresponding to 1000 samples and then averaged. Since in this work the system cannot be characterized for all possible configurations, only one combination of excitation current and frequency was used. The excitation signal was chosen to be I=1 mA with a frequency of f=100 kHz.

In Fig. 6 the true impedances as well as the measured values are plotted. To analyze if the actual magnitude or phase of the impedance under test has an influence to the systematic error, both were varied. Table I summarizes the obtained values of the absolute and relative systematic measurement errors under the variation of the magnitude and the phase.

The relative magnitude errors were measured to be lower than 2 ‰ and the phase errors are below 0.2° in the range from 7.3 $\Omega$ to 373.6 $\Omega$. These values are in the range of the tolerances of the used test impedances. The observed systematic errors are sufficient for the intended application.

### B. Statistical Errors

Especially for detecting the expected impedance changes reliably, it is important to determine the statistical errors of the system. For this purpose, the known test impedances are measured similarly as for the systematic error measurement but with a focus on statistical errors. In Fig. 7, an exemplary measurement with a duration of 3 s of one of these impedances is shown.

For calculating the standard deviations (STD) of the magnitudes and phases, measurements with durations of 1 s are used.

In Table II, the obtained values are again shown under variation of the impedance's magnitude and phase. Since it is assumed that muscle contractions cause impedance changes in $\Omega$-ranges [18], the standard deviation of maximum 202 m$\Omega$ is sufficiently small. The standard deviation of the phase measurement for all setups is lower than 0.03°. The results in the table II also show, that the standard deviations of the magnitude as well as of the phase do not significantly depend on the absolute values of the analyzed impedance.

### C. Long-Term Drift

To estimate the long-term drift of the device, a measurement of a 7.3 $\Omega$ test impedance with a phase of -25.7° was performed for a duration of 20 minutes. The absolute drift during this time period was determined to be lower than 0.5 m$\Omega$ according to

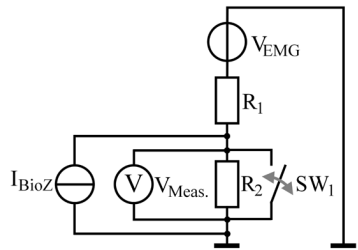

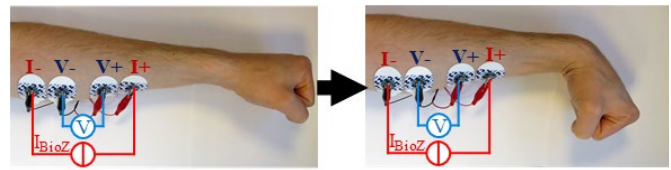

Fig. 8. Setup to generate simultaneous impedance and EMG signal steps to analyze the systems' synchronicity. In this figure each, the $I_{BioZ}$ and $V_{Meas.}$ represent the four channels of the four measurement modules.

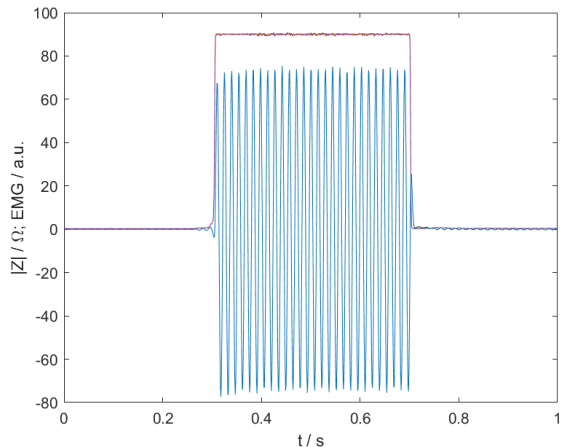

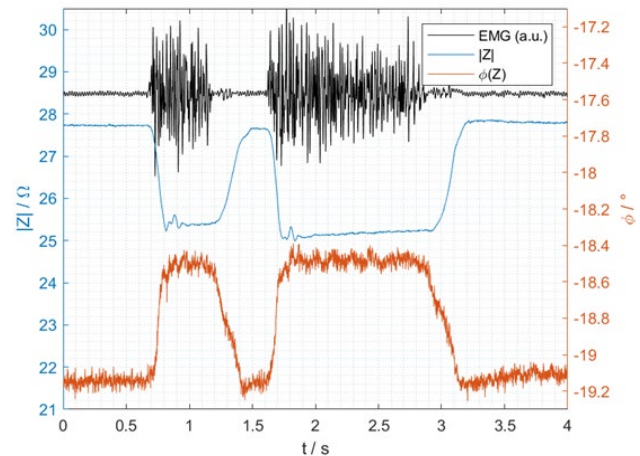

Fig. 9. Acquired step response of the four bioimpedance measurement channels as well as the four EMG channels. Because of the high synchronicity and precision, the plot seems to display just two signals. The time delay between the EMG channels and the bioimpedance channels was determined to be lower than 10 ms.

57 ppm. The phase drifted during this time span about 0.007°.

### D. Synchronicity

The synchronicity of all impedance and EMG channels is mandatory to be lower than 100 ms to enable prostheses or other devices to react within a time-span which might be acceptable for the user. Additionally, the synchronicity avoids misinterpretations of movements if the signals change very quickly.

To generate a signal step and to measure it with all channels simultaneously, a setup as shown in Fig. 8 was used. In this setup, the resistor $R_2$ represents a bioimpedance, which is measured by the four measurement modules. In this figure $I_{BioZ}$ represents the four current sources which apply currents of 100 µA each through $R_2$. To enable the separation of the channels, different excitation frequencies were used ($f_1$=91 kHz, $f_2$=100 kHz, $f_3$=111 kHz, $f_4$=125 kHz). Accordingly, $V_{Meas.}$ represents the voltage measurement sections of all the four modules. To generate an artificial EMG signal, $V_{EMG}$ was realized by using a signal generator (SDG1050, Siglent, Helmond, NL), which generated an AC voltage of 1 $V_{pp}$ with a frequency of 70 Hz. $R_1$ realized in combination with $R_2$ a 10:1 voltage divider to decrease the artificial EMG voltages to suitable values. To generate signal steps, the switch $SW_1$ was opened and closed alternately.

In Fig. 9, the resulting measured signals of the impedance magnitudes and artificial EMG voltages are shown. The plot displays all eight signals. The switch $SW_1$ was opened after about 0.3 s for a duration of about 0.4 s and closed again. It can be seen, that the four EMG signals as well as the four bioimpedance signals are that synchronous, that the plots are one above the other and it looks as if there are just two signals plotted. Also the time delay between the measured EMG signal step and the bioimpedance step is below 10 ms and therefore sufficiently short.

Fig. 10. Measurement setup to acquire the muscle contraction during wrist flexions and a plot of the measured raw data of the EMG signal and the complex bioimpedance. The muscle contraction was performed two times. In comparison to the EMG signal, it can be seen, that especially the beginning of the contraction can also easily be detected by performing bioimpedance measurements.

## V. MEASUREMENTS

Measurements were performed on a human subject to demonstrate the usability of the measurement system. After comparing real EMG and EIM data with each other, the robustness of both signals towards mechanical interferences is analyzed. Finally, a set of typical hand gestures is measured by using all available measurement channels.

### A. Comparison between EMG and BioZ-Signals

Previous publications focus either on the EMG or on the EIM signals [13], [15], [21], [22]. The relationship between both the signals, especially the timing behavior has not been investigated reliably, so far. Therefore, a muscle contraction of the forearm, as shown on top of Fig. 10 was performed. To acquire the signals directly above the musculus flexor carpi radialis, four Ag/AgCl gel electrodes (Kendall H92SG, Covidien, Dublin, IE) were placed on the forearm and connected to a measurement module as depicted in the figure. The distance between the voltage electrodes was measured to be 3 cm.

The configured bioimpedance excitation current was 100 µA with a frequency of 100 kHz and the chosen gain was G=25.

Within the measurement time span of four seconds, the muscle contraction was performed two times for durations of about 0.5 s and 1.3 s.

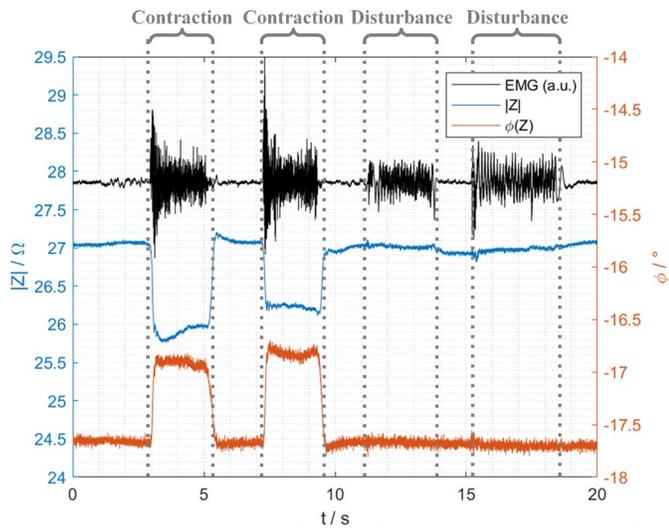

Fig. 11. Measurement results of two performed muscle contraction of the forearm flexor muscle and two artificially applied mechanical disturbances to the positive voltage electrode. It can be seen that the applied vibrations affect the EMG signal significantly, whereas the bioimpedance magnitude and phase do almost not change during these time periods.

The acquired unfiltered data of the EMG signal (black), the bioimpedance magnitude (blue) and the bioimpedance phase (red) are shown in the plot in Fig. 10. For comparison purposes, the EMG signal is plotted in arbitrary units.

Both contractions can clearly be seen in the morphology of the EMG signal. It is also distinguishable, that during these contractions, the magnitude of the bioimpedance decreases from about 27.7 Ω to approximately 25.4 Ω. During this time period, the phase shift increases by about 0.7°. It can be seen that the slopes of the impedance signals begin simultaneously with the EMG signals. After approximately 100 ms the bioimpedance values reach a constant state, which indicates the end of the movement and therefore the end of geometrical muscle changes. After contractions, the bioimpedance signals need more time to reach to original values, which might be caused by the different signal origins.

This measurement demonstrates that the acquisition of the complex bioimpedance, simultaneously to EMG signals, provides additional useful information.

### B. Robustness towards Mechanical Disturbances

The most promising advantage of the additional bioimpedance recording is its robustness towards electrode motion artifacts. By only evaluating the occurring voltage signals of the known excitation frequency, interferences of all other signals with different frequencies can be eliminated.

To evaluate this beneficial effect, the measurement of the section above was repeated under the same conditions. In addition to contracting the forearm muscle two times, mechanical disturbances were applied to the positive voltage electrode to generate changes in the half-cell voltage. These minuscule vibrations were generated by slipping the clamping mechanism of a cordless screwdriver on the electrode with an intensity which produces disturbances in the range of the actual EMG signal.

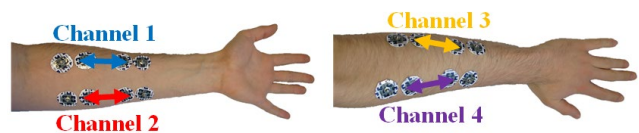

Fig. 12. Positioning of the 16 Ag/AgCl gel electrodes for performing measurements with four measurement modules. The distance between related voltage electrodes is about 4 cm.

In Fig. 11 the acquired three raw signals are displayed. During the two performed muscle contractions, the signals show up as before. During the two additional periods (11-14 s; 15-18.5 s), the previously described external force was applied to the electrodes. While the EMG trace shows strong interferences with amplitudes similar to the former EMG signals, the EIM magnitude |Z| is only slightly affected at the onset of the mechanical disturbances. The EIM phase however remains not influenced.

### C. 4-Channel-Measurement for Prosthesis Control

In many applications, for instance prostheses control, it is desirable to analyze several muscle groups simultaneously. Especially, the geometrical depth of the impedance measurement needs to be considered. If thus particular muscle regions cannot be selected by a single set of electrodes reliably the simultaneous use of several electrode sets could be useful.

To demonstrate the system's capability to distinguish more complex activations of muscle groups, measurements utilizing all four available measurement modules were performed on a human forearm. The chosen positions of the Ag/AgCl gel electrodes are shown in Fig. 12. The excitation currents of the modules were 200 µA and to separate the channels from each other, different excitations frequencies were chosen ($f_1$=91 kHz, $f_2$=100 kHz, $f_3$=111 kHz, $f_4$=125 kHz).

To verify whether the bioimpedance signal provides additional information or if it is just redundant to the EMG signal, several typical wrist movements were performed by the subject. The chosen five movements are shown on the top of Fig. 13. Under the corresponding photographs of the movements, the acquired data of the four EMG, |Z| and φ(Z) channels are shown. Only the EMG signal is filtered by a zero-phase notch filter to reduce the 50 Hz interferences. No further digital filtering is performed to the signals. For a more convenient perception, offsets are added to the EMG signals. The magnitude and phase plots present the occurring changes in percent and in degrees, respectively.

Every performed muscle contraction was started after 1 s for a duration of about 2 s. These periods can easily be distinguished in all signal plots.

Comparing the five impedance magnitude plots leads to the assumption that the channel decoupling in this measurement setup is sufficient. This is especially visible when focusing for example on the signals of the channels CH2 (red) and CH3 (yellow) with adjacent electrodes and similar frequencies of the first (WF) magnitude plot. Even when both sets of electrodes and thus the corresponding bioimpedances are close to each other, the behavior of the signals is completely different.

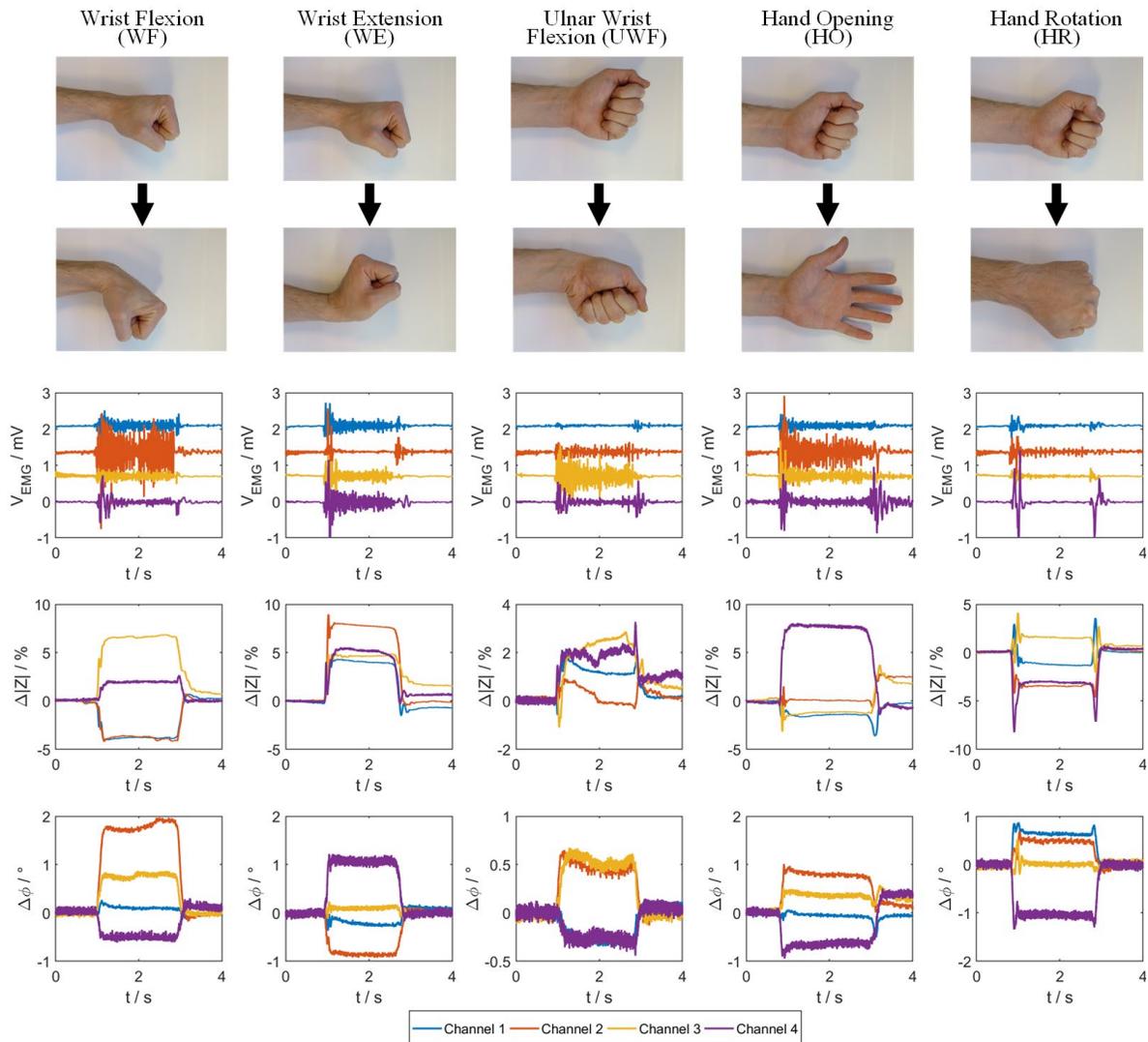

Fig. 13. Measurement result of five typical hand gestures, using the four measurement modules simultaneously. Each muscle contraction was performed for a duration of about 2 s. In the plots below the corresponding photographs, the EMG signals as well as the bioimpedance magnitudes and phases are shown. The EMG data was preprocessed by a digital 50 Hz notch filter, whereas the impedance signals have not been filtered digitally. Comparing the resulting plots leads to the conclusion that the magnitude and phase signals of the bioimpedance provide additional useful information about muscle contractions. Additionally, hand gestures which are difficult to detect via EMG measurements, for example the hand rotation (HR), can be recognized via bioimpedance measurements reliably.

To check if the information of the EMG data and the corresponding impedance magnitudes is same, the plots of the wrist flexion (WF) and extension (WE) can be compared. It can be seen that the EMG behavior of CH1 (blue) during both the movements has a similar behavior. In contrast, the corresponding impedance magnitude signals of CH1 differ completely from each other. The same effect can be found when comparing the EMG data with the impedance phase signal. For example, the EMG signals of CH4 look similar during the WE and the HO motion, whereas the behavior of the corresponding impedance phases is opposite during both the muscle contractions. Finally, comparing the impedance magnitude and phase plots in the same way, leads to the result, that acquiring EMG, bioimpedance magnitude and phase provides three independent fully redundant sets of information per measurement channel.

### D. Measurement of Isometric Muscle Contractions

The previously presented measurements were performed during actual movements. In some applications it might also be interesting to detect isometric muscle contractions without significant geometrical changes within the tissue under test. To give an impression about the occurring bioimpedance signals during these kind of muscle contractions, the previously shown measurement was repeated for the wrist flexion and extension. Both contractions were performed against solid plates, as shown at the top of figure 14, to avoid changes in the angle between the forearm and the hand. The occurring forces $F_{iso}$ during the contractions were measured to be 20±3 N. Both contractions were performed two times for a duration of about 2 s.

In the corresponding EMG plots in figure 14, it can be seen that the signal behavior is similar to the previously presented results of the non-isometric wrist flexion and extension. The isometric contractions do also cause significant changes of the bioimpedance magnitude and phase signals. In comparison to the results in figure 13, the relative changes of the bioimpedance magnitudes and the phase shifts are lower. Some signals, such as CH3, have a very similar behavior during isometric and non-isometric contractions. However, the signal of CH2 behaves very different in both cases. This might be caused by the different kinds of geometrical changes within the tissue, but has to be analyzed more detailed in the future.

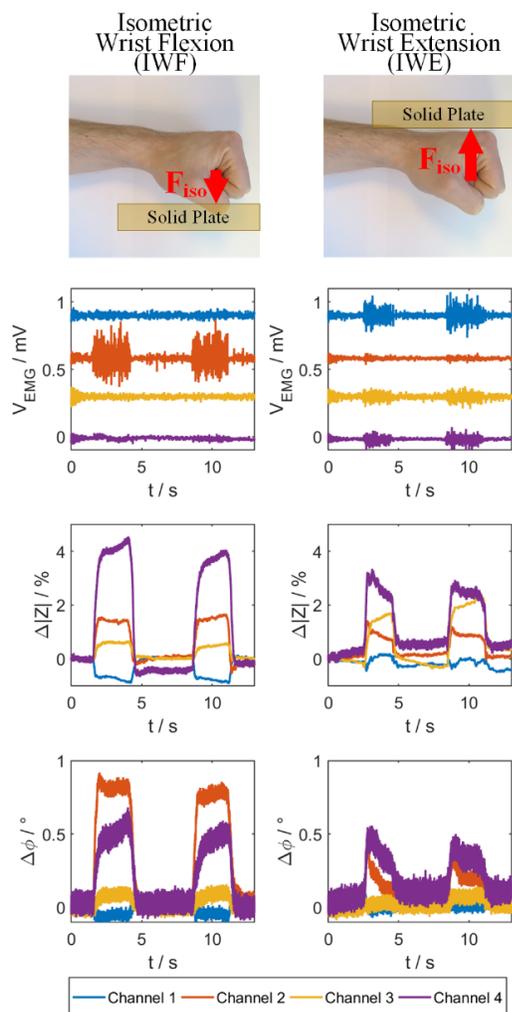

Fig. 14. Measurement of two typical hand gestures. To enable isometric muscle contractions, the movements were performed against solid plates with a force of approximately 20 N.

## VI. Discussion

The EMG and the EIM, though both deployed to detect muscle activations, do differ with respect to their physiologic origins. EMG represents the bioelectrical signal of the muscle contraction caused by the muscle cells' activation potentials. The impedance signal however, is induced by the geometrical changes of the muscle. Either during active contraction or during passive expansion. Therefore, both the signals do not strictly depend on each other. Since there is usually a time delay between muscle activation and its geometrical reaction [23], the bioimpedance signal is delayed against the EMG signal. This delay is visible in the exemplary measurement shown in Fig. 10. It occurs in a range of tens of milliseconds and it depends on the particular application, if such delays are acceptable.

The measurement, in Fig. 11 has demonstrated that the sensitivity regarding mechanical disturbances of the voltage electrodes is much lower in bioimpedance measurements than in EMG signals. Nevertheless, the conditions of the electrode skin contacts in EIM applications is important as well. High interface impedances of the current electrodes lead to high voltage drops over the electrode skin impedances. This issue has a significant impact on the design of the current source as well as on the voltage measurement circuit [24]. The usage of dry electrodes, as preferred in prostheses, with commonly high electrode skin impedances aggravates this issue [25], [26]. But lower excitation currents or higher excitation frequencies can help to keep the voltage drop over the mainly capacitive impedances under a tolerable limit.

The measurement results shown in Fig. 13 are very promising. They support that the additional bioimpedance measurement triples the amount of usable information. In this particular example, 12 signals are available to detect the muscle contractions in the forearm. Even in situations in which the EMG signals are poor, it seems that the bioimpedance measurements provide useful information. The information of both kinds of signals can be combined by applying machine learning algorithms. The results in figure 14 show that the bioimpedance magnitude and phase change also during isometric contractions.

It has to be noted, that all measurements in this work were performed in a laboratory environment. Further disturbances such as the mechanical interactions between the electrodes and the muscles were not investigated. It is conceivable that changes of the limb position can lead to impedance changes as well. Exhaustive studies on subjects are mandatory to analyze the influence of the electrode-skin interfaces and the electrode positions on the measurement results.

Furthermore, the anisotropic behavior of the muscles' impedance has not been analyzed in this work. This behavior could also be a useful information to differentiate between motion artifacts and actual muscle contractions.

To assess the actual benefit of the additional bioimpedance signals for different prostheses situations as well as for other applications, more studies with extended subject sets and typical environmental noise situations are envisaged for the future.

## VII. Conclusion

In this work we presented an approach to acquire EMG and EIM information about muscle contractions from several positions of a subject, simultaneously.

A modular multi-channel measurement device was developed and characterized. In addition to an ordinary EMG system, this device is capable of acquiring bioimpedance magnitude changes in m$\Omega$ ranges and phase changes smaller than 0.05°.

First measurements from a human subject have shown the differences between EMG and EIM signals with a focus on their timing relationships. Additionally, the robustness of bioimpedance measurements towards mechanical disturbances to the electrodes was demonstrated.

Finally, the combination of four measurement modules was used to analyze typical hand gestures by combining EMG and bioimpedance measurements. The results demonstrated the usefulness of acquiring the bioimpedance to detect muscle motions, especially for prosthesis controls. But many other human-machine interface applications are conceivable as well.

Even when the proposed results are very promising, further studies on the influence of the electrodes' behavior and positions have to be performed.

In the future, existing algorithms have to be adapted to combine the signals and to convert them into a movement information, automatically.